\documentclass[superscriptaddress,twocolumn,notitlepagem,longbibliography,prx]{revtex4-1}

\usepackage{graphicx}
\usepackage{float} 
\usepackage{subfigure} 
\usepackage{dcolumn}
\usepackage{bm}
\usepackage{amsthm}
\usepackage{amsmath}
\usepackage{amssymb}
\usepackage{color}
\usepackage{algpseudocode}
\usepackage[normalem]{ulem}
\usepackage{natbib}
\usepackage{hyperref}
\definecolor{Red}{rgb}{1,0,0}
\def\vec#1{{\bm #1}}

\def\ket#1{| #1 \rangle}
\def\bra#1{\langle #1 |}

\usepackage[normalem]{ulem}

\begin{document}

\title{Ensemble-learning error mitigation for variational quantum shallow-circuit classifiers}

\author{Qingyu Li}
\affiliation{Institute of Fundamental and Frontier Sciences, University of Electronic Science and Technology of China, Chengdu, 610051, China}

\author{Yuhan Huang}
\affiliation{The Department of Electronic and Computer Engineering, The Hong Kong University of Science and Technology, 999077, Hong Kong}

\author{Xiaokai Hou}
\affiliation{Institute of Fundamental and Frontier Sciences, University of Electronic Science and Technology of China, Chengdu, 610051, China}

\author{Ying Li}
\email{yli@gscaep.ac.cn}
\affiliation{Graduate School of China Academy of Engineering Physics, Beijing 100193, China}

\author{Xiaoting Wang}
\email{xiaoting@uestc.edu.cn}
\affiliation{Institute of Fundamental and Frontier Sciences, University of Electronic Science and Technology of China, Chengdu, 610051, China}

\author{Abolfazl Bayat}
\email{abolfazl.bayat@uestc.edu.cn}
\affiliation{Institute of Fundamental and Frontier Sciences, University of Electronic Science and Technology of China, Chengdu, 610051, China}

\begin{abstract}
Classification is one of the main applications of supervised learning. Recent advancement in developing quantum computers has opened a new possibility for machine learning on such machines. Due to the noisy performance of near-term quantum computers, error mitigation techniques are essential for extracting meaningful data from noisy raw experimental measurements. Here, we propose two ensemble-learning error mitigation methods, namely bootstrap aggregating and adaptive boosting, which can significantly enhance the performance of variational quantum classifiers for both classical and quantum datasets. The idea is to combine several weak classifiers, each implemented on a shallow noisy quantum circuit, to make a strong one with high accuracy. While both of our protocols substantially outperform error-mitigated primitive classifiers, the adaptive boosting shows better performance than the bootstrap aggregating. The protocols have been exemplified for classical handwriting digits as well as quantum phase discrimination of a symmetry-protected topological Hamiltonian, in which we observe a significant improvement in accuracy. Our ensemble-learning methods provide a systematic way of utilising shallow circuits to solve complex classification problems. 
\end{abstract}

\maketitle

\section{INTRODUCTION}
Machine learning, as a method in which computers learn patterns within data, has revolutionized almost all aspects of our lives~\cite{mitchell_machine_1997}. Classification algorithms are among the most important types of machine learning tasks with a wide range of applications in finance, business, industry, marketing, and scientific research~\cite{Review_classification,zhang_review_2014}. In these algorithms, all data are divided into a few discrete classes that contain elements with certain common features. So far, numerous classification algorithms have been developed, such as logistic regression~\cite{tolles_logistic_2016}, decision trees~\cite{quinlan_induction_1986}, $k$-nearest neighbors~\cite{1053964}, support vector machines~\cite{cortes_support-vector_1995}, and neural network classifiers~\cite{897072}. 
The sophistication that big data brings to the training process may decrease the accuracy of algorithms. To overcome this, one can adopt ensemble-learning methods in which several classifiers are combined to make a stronger one with higher prediction accuracy. The most prominent ensemble-learning methods are Bootstrap Aggregating (Bagging)~\cite{breiman_bagging_1996} and Adaptive Boosting (AdaBoost)~\cite{freund_desicion-theoretic_1995,hastie_multi-class_2009} which have been developed in classical machine learning context.  

Quantum computers are rapidly emerging in various physical platforms, including superconducting qubits~\cite{barends_digitized_2016,hensgens_quantum_2017,arute_quantum_2019,google_ai_quantum_and_collaborators_hartree-fock_2020,PhysRevLett.127.180501,bravyi2022future}, ion-traps~\cite{hempel_quantum_2018,kokail2019self,ringbauer2022universal,noel2022measurement,monroe2021programmable}, optical lattices~\cite{schreiber_observation_2015,gross_quantum_2017,sompet2022realizing}, Rydberg atoms~\cite{saffman_quantum_2016} and photonic chips~\cite{aspuru-guzik_photonic_2012,doi:10.1126/science.abe8770,dai2022topologically}. 
They can push our computational power well beyond the capability of existing classical computers~\cite{arute_quantum_2019, PhysRevLett.127.180501, doi:10.1126/science.abe8770,daley2022practical}.  
Indeed, several classification algorithms have been generalized to be adopted on quantum computers, including distance-based quantum classifier~\cite{Schuld_2017}, quantum support vector machine~\cite{rebentrost_quantum_2014,li_experimental_2015}, quantum $k$-nearest neighbor algorithm~\cite{lloyd_quantum_2013, wiebe_quantum_2014}, quantum decision tree classifiers~\cite{lu_quantum_2014,PhysRevA.58.915}, and quantum neural networks~\cite{grant_hierarchical_2018,schuld_circuit-centric_2020,classification_Farhi,cong_quantum_2019,PhysRevResearch.1.033063,Romero_2017,zoufal2019quantum,hou2022duplication,liu2021hybrid}. 
However, current Noisy Intermediate-Scale Quantum~(NISQ) computers are far away from achieving fault-tolerant quantum computing~\cite{bharti_noisy_2022} and thus most of these algorithms cannot be realized on such noisy hardware. 
In fact, outperforming classical computers with NISQ devices is still a hotly debated subjected~\cite{RevModPhys.94.015004}. 

Variational quantum algorithms~(VQAs)~\cite{cerezo_variational_2021} are the most promising approach for achieving quantum advantage on NISQ computers. In these algorithms, the complexity is divided between a quantum circuit and a classical computer, allowing a complex task to be achieved using a shallow quantum circuit. 
So far, VQAs have been exploited to solve a wide range of problems, including eigenvalue solvers~\cite{peruzzo_variational_2014,kandala_hardware-efficient_2017,higgott_variational_2019,PhysRevLett.122.140504,Lyu2020accelerated,lyu2022symmetry}, quantum neural networks~\cite{biamonte2017quantum,PhysRevA.98.032309,cong_quantum_2019}, quantum adversarial machine learning~\cite{PhysRevResearch.2.033212,PhysRevResearch.3.023153,PhysRevA.101.062331,ren2022experimental}, quantum approximate optimization algorithms~\cite{farhi_quantum_2014}, linear equation solvers~\cite{bravo-prieto_variational_2020,xu_variational_2021,Huang_2021} and quantum sensing~\cite{beckey_variational_2022,kaubruegger_variational_2019,meyer_variational_2021}.
Variational Quantum Classification (VQC) algorithms, as typical VQAs, have also been developed to solve classification problems on NISQ computers~\cite{cong_quantum_2019,grant_hierarchical_2018,classification_Farhi,banchi_robust_2022,PhysRevA.101.032308,deng_quantum_2021,gong_universal_2021,PhysRevResearch.3.023153,classification_Farhi,PhysRevA.102.012415,ren2022experimental}, with some of them being experimentally demonstrated~\cite{herrmann_realizing_2022,havlicek_supervised_2019,arute_quantum_2019,ren2022experimental}.  
Nonetheless, the imperfect nature of NISQ computers restricts the achievable accuracy of VQCs. 
In the absence of error correction, error mitigation techniques~\cite{PhysRevX.7.021050,temme_error_2017,kandala_error_2019,endo_hybrid_2021,qin2022overview,cai2022quantum} have been developed to extract noise-free values from noisy experimental data. Zero-Noise Extrapolation (ZNE) is one of the most practical error mitigation methods, in which one can measure the desired observable at different levels of noise. Then by extrapolation, one can estimate the noise-free value through a proper fitting function~\cite{PhysRevX.7.021050,temme_error_2017,kandala_error_2019}. Developing novel error-mitigation techniques is essential for achieving practical quantum advantage using near-term quantum computers. Indeed, one open problem is whether one can combine successful ensemble learning techniques, developed in classical machine learning, with VQCs to enhance the precision of classifiers for both classical and quantum datasets. 

In this paper, we develop two ensemble-learning error mitigation techniques, namely Bagging and AdaBoost, for VQCs to combine a few weak quantum classifiers and make a strong one with enhanced accuracy. This allows to use of shallow circuits for each of the classifiers and improves noise resilience. We find that both of the proposed protocols significantly outperform the ZNE method in classification tasks. 
In the two protocols, the AdaBoost shows stronger performance, namely higher accuracy and more noise resilience, than the Bagging.
Since our ensemble learning algorithms can achieve accurate classification with only shallow circuits, they are NISQ-friendly and feasible for applications using existing quantum-computing technologies. 
This makes them very distinct from other quantum versions of ensemble-learning algorithms~\cite{schuld_quantum_2018,wang_quantum_2020,arunachalam_quantum_2020,izdebski_improved_2020,macaluso2020quantum} which utilize deep-circuit quantum subroutines, such as quantum phase estimation~\cite{kitaev_quantum_1995}, quantum means estimation~\cite{10.1145/301250.301349,brassard_optimal_2011} and Grover search~\cite{10.1145/237814.237866} algorithms, to speed up the training process and reduce the sample complexity. 

We noted that a plain ensemble learning method, namely plurality voting, had been applied to improve the performance of VQCs~\cite{qin2022improving}. 
However, the plurality voting method is mainly used to reduce variance and improve the robustness of the model to data, but can not significantly improve the accuracy of the model. 

\section{VARIATIONAL QUANTUM CLASSIFIERS}
\subsection{Classification Problems}
Classification tasks are types of supervised machine learning problems in which the goal is to predict a discrete class label $y$, for a given unknown data $\vec{x}$. 
In general, the classifier is trained by a labeled training dataset with $M_\text{s}$ samples, $\hat{\mathcal{D}}{=}\{(\vec{x}_i,y_i)\}_{i=1}^{M_\text{s}}$, where $\vec{x}_i{=}[x_{i1},\cdots,x_{iN_\text{f}}]^\mathsf{T}$ represents an input vector with $N_\text{f}$ features and $y_i$ is the corresponding class label which takes $K_\text{c}$ different values (i.e. $y_i\in\{1,\cdots,K_\text{c}\}$).
After training, the classifier can be described as a map $\hat{y}=f(\vec{x})$ where $\hat{y}$ is the predicted label for a given input $\vec{x}$. 
For a good classifier, we expect that $y=\hat{y}$, namely predicting the correct class label. 
In reality, our prediction might be wrong for some inputs, nonetheless, the objective is to keep the ratio of wrong predictions  as small as possible.

\begin{figure*}
  \centering
  \includegraphics[width=0.85\textwidth]{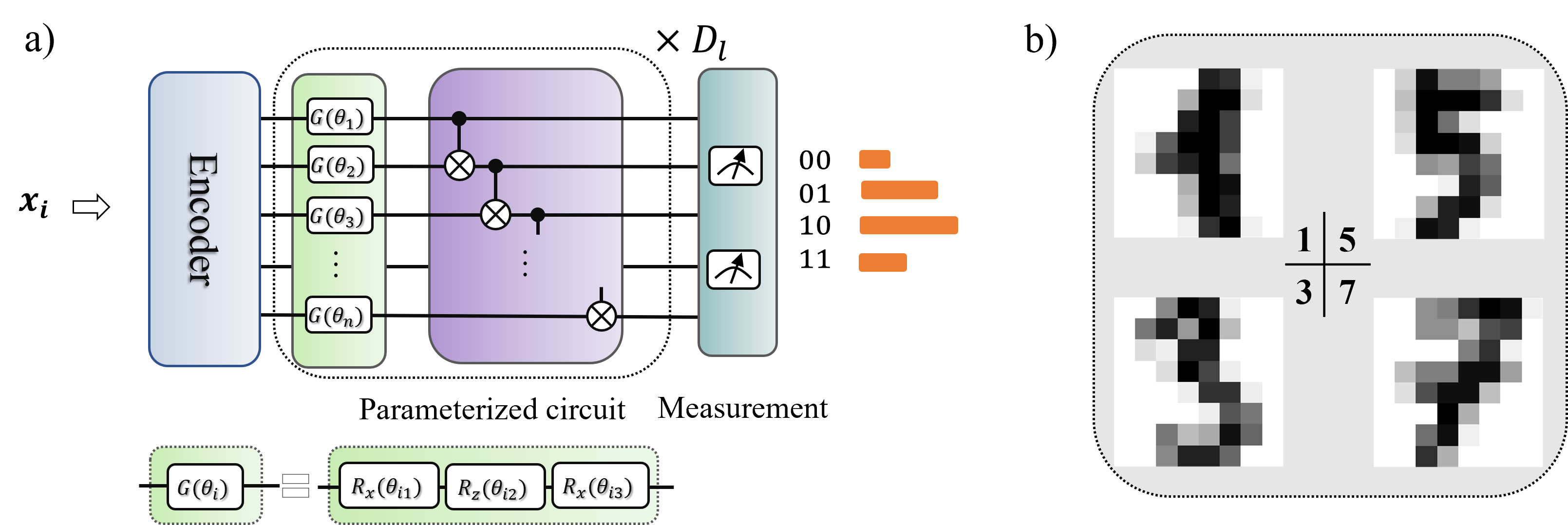}
  \caption[]{\textbf{Circuit design and the classical dataset.} (a) The quantum circuit used in our ensemble-learning VQC protocols contains three different parts, namely encoder, parameterized circuit and measurement. The encoder part transforms classical input data into a quantum state. While in the paper, we use amplitude encoding the protocol works equally well for rotation encoding too. For quantum datasets, the encoder part is not needed. The prepared quantum states are fed into a parameterized circuit in  which each qubit first undergoes a local rotation $G^{(q)}(\vec{\theta}_{q})=R_x^{(q)}(\theta_{q1})R_z^{(q)}(\theta_{q2})R_x^{(q)}(\theta_{q3})$, shown in the lower panel, and then a series of two-qubit controlled-not gates act on nearest neighbor qubits.   The whole parameterized circuit  is repeated $D_l$ times.
  Then the quantum measurement is applied to a few qubits which depend on the number of classes, for obtaining the probabilities of different labels. The label with the largest probability is chosen as the final prediction label. 
  (b) Four typical images of the MNIST dataset which show handwriting digits ${1,3,5,7}$. Each image contains $8\times 8$ pixels which are reshaped as a normalized 64-dimensional vector $\vec{x}_i$ as the input data. The dataset has 1541 training samples and 726 test samples with these four digits.}
  \label{Fig.circuits}
\end{figure*}

Recent advancements in developing quantum computers have opened a new territory for exploiting such machines for solving classification problems. In this case, apart from solving conventional classification problems, which deal with classical datasets, one can also consider quantum datasets $\hat{\mathcal{D}}{=}\{(\ket{\vec{x}_i},y_i)\}_{i=1}^{M_\text{s}}$, where $\ket{\vec{x}_i}$ is a quantum state to represent the input features and $y_i$ is the corresponding class label  which takes $K_\text{c}$ different values.  
The inherent nature of quantum datasets $\hat{\mathcal{D}}$ justifies the use of a quantum classifier as no classical counterpart can be used for such data. The situation is, however, very different for classical datasets as it is still an open question whether the full capacity of quantum computers can be exploited for the classification of classical data. 

In general, one constructs a classifier $f(\vec{x})$ by training it on dataset $\hat{\mathcal{D}}$, which can be either classical or quantum.
We denote the accuracy of $f(\vec{x})$ as $1{-}e$ where $e$ is error rate as
\begin{equation}
  e=\frac{1}{M_\text{s}}\sum_{i=1}^{M_\text{s}} \mathbb{I}(f(\vec{x}_i)\neq y_i),
  \label{error_rate}
\end{equation}
where $\mathbb{I}(\cdot)$ is the Indicator function with $\mathbb{I}(\cdot){=}1$ for $(\cdot)$ being True and $\mathbb{I}(\cdot){=}0$ otherwise. A random guess determines the class label correctly with a probability $1/K_\text{c}$ and thus its error rate, given in Eq.~(\ref{error_rate}), would be $e{=}(K_\text{c}{-}1)/K_\text{c}$. A given classifier $f(\vec{x})$ if called strong if $e{\sim} 0$ and is called weak if $e{\sim} (K_\text{c}{-}1)/K_\text{c}$. 

\subsection{A General Implementation of Variational Quantum Classifiers}
Variational Quantum Algorithms (VQAs) are the most promising approach for achieving quantum advantage on NISQ computers. In these algorithms, the complexity is divided between a quantum circuit and a classical optimizer. Therefore, even a shallow quantum circuit might be sufficient to achieve a complex task.
Recently, VQAs have also been used for developing quantum classifiers for both classical~\cite{grant_hierarchical_2018,schuld_circuit-centric_2020,classification_Farhi} and quantum~\cite{cong_quantum_2019,classification_Farhi,PhysRevA.102.012415} datasets. 
Nonetheless, in the NISQ era, developing new techniques for mitigating the effect of noise is essential for scaling up the classification algorithms to deal with more complex datasets which normally demand larger numbers of qubits and deeper circuit depths.   

In this work, we focus on Variational Quantum Classifiers (VQC). The VQC circuits contain three parts: encoding circuit, parameterized circuit, and measurement. The schematic representation of the circuit is shown in Fig.~\ref{Fig.circuits}(a).
For quantum datasets, the encoding circuit is not needed as the data can directly be fed into the parameterized circuit. For classical datasets, however, the input data  $\vec{x}_i$ has to be encoded into a quantum state $\ket{\Vec{x}_i}$. Amplitude encoding is the most efficient way for converting classical data into a quantum state with an exponential advantage through mapping $N_\text{f}$ features into $N_\text{q}{=}\left \lceil  \log_2(N_\text{f})\right \rceil$ qubits as
\begin{equation}
    \Vec{x_i}\to\ket{\vec{x}_i}=\frac{1}{||\vec{x}_i||}\sum_{j=1}^{N_\text{f}}x_{ij}\ket{j},
\end{equation}
where $||\vec{x}_i|| {=} \sqrt{\vec{x}_i^\mathsf{T} \vec{x}_i}$ is the norm of $\vec{x}_i$ and $\ket{j}$ is a quantum state of $N_\text{q}$ qubits with binary representation of $j$ in the computational basis. This encoding is assumed to be done through a Quantum Random Access Memory (QRAM) module~\cite{casares_circuit_2020,jiang_experimental_2019,park_circuit-based_2019,giovannetti_quantum_2008}. 
It is worth emphasizing that our protocol does not depend on any specific encoding method and can easily be generalized to other encoders, such as rotation encoding~\cite{schuld_effect_2021,PhysRevA.102.032420,PhysRevA.98.032309}. Therefore, for the sake of brevity, we only focus on amplitude encoding.  
The output of the encoder is fed into a parameterized circuit that contains several layers. Each layer of the parameterized circuit starts with a series of local rotations $\prod_{q} G^{(q)}(\vec{\theta}_q)$ acting on all qubits with
\begin{equation}
G^{(q)}(\vec{\theta}_{q})=R_x^{(q)}(\theta_{q1})R_z^{(q)}(\theta_{q2})R_x^{(q)}(\theta_{q3}),
\end{equation}
where $\vec{\theta}_q{=}[\theta_{q1},\theta_{q2},\theta_{q3}]^\mathsf{T}$, $R_\alpha^{(q)}(\theta){=}e^{-i\theta\sigma_\alpha^{(q)}/2}$ (for $\alpha{=}x$ or $z$) and $\sigma_\alpha^{(q)}$ is the Pauli operator $\alpha$ acting on qubit $q$. The single qubit rotations are followed by a series of two-qubit controlled-not gates $\prod_q U_{CX}^{(q,q+1)}$ with
\begin{equation}
U_{CX}^{(q,q+1)}=\ket{0}\bra{0}^{(q)}\otimes I^{(q+1)} + \ket{1}\bra{1}^{(q)}\otimes \sigma_x^{(q+1)},
\end{equation}
where $I^{(q)}$ represents identity acting on qubit $q$.
Therefore, the action of the parameterized circuit with $D_l$ layers on $N_\text{q}$ qubits can be described by a unitary operator of the form
\begin{equation}
   U(\vec{\theta}) = \prod_{d=1}^{D_l}\left(\prod_{q=1}^{N_\text{q}-1} U_{CX}^{(q,q+1)} \prod_{q=1}^{N_\text{q}} G^{(q)}(\vec{\theta}_{dq}) \right).
\end{equation}
The schematic of the circuit is shown in Fig.~\ref{Fig.circuits}(a).
The output of the circuit is  given by $U(\vec{\theta})\ket{\vec{x}_i}$. By measuring the last few qubits of the circuit one can determine the class label of the input $\ket{\vec{x}_i}$. 
In fact, the number of qubits that are measured is determined by $\left \lceil  \log_2(K_\text{c})  \right \rceil$.
The measurement outcomes can be described by projectors $\{ \Pi_k \}^{K_\text{c}}$, where $K_\text{c}$ is the  number of classes.  For instance, for binary classification (i.e. $K_\text{c}{=}2$), only the last qubit is measured and the projectors are given by $\Pi_0{=}\ket{0}\bra{0}$ and $\Pi_1{=}\ket{1}\bra{1}$, acting on qubit $N_q$. 
Similarly, for a four-class problem, one has to measure the last two qubits (namely qubits $N_\text{q}{-}1$ and $N_\text{q}$), and the classes are determined by projectors
$\Pi_{k}\in \{ \ket{00}\bra{00}, \ket{01}\bra{01}, \ket{10}\bra{10}, \ket{11}\bra{11} \}$, which act on qubits $N_\text{q}{-}1$ and $N_\text{q}$.
The probabilities of measurement outcomes are considered as the probabilities of obtaining each class label $\hat{y}_i$, as $\vec{p}_i{=}[p_{i1},\cdots,p_{iK_\text{c}}]^\mathsf{T}$,
where 
\begin{equation}
        p_{ik} =\bra{\vec{x}_i}U^\dagger(\vec{\theta}) \Pi_{k}U(\vec{\theta})\ket{\vec{x}_i}.
\end{equation}
The label $\hat{y}_i$ is determined by the class $k$ whose  probability $p_{ik}$ is maximum, namely $\hat{y}_i{=}\text{arg}\max_{k}p_{ik}$. The schematic of the procedure is shown in Fig.~\ref{Fig.circuits}(a).
The performance of VQC is evaluated by a loss function described by cross-entropy  
\begin{equation}
  \mathcal{L}(\vec{\theta})=-\sum_{i=1}^{M_\text{s}}\Vec{y}_i^\mathsf{T}\log(\Vec{p}_i),
\end{equation}
where $\vec{y}_i{=}[y_{i1},\cdots,y_{iK_\text{c}}]^\mathsf{T}$ is the one-hot encoding of the true class label $y_i$ with only one of the elements $y_{ik}$, which is the right class label, is 1 and the rest are 0.
By using Adam optimizer~\cite{kingma_adam_2017}, which is a gradient-based method, one can iteratively update $\vec{\theta}$ in order to minimize the loss function. More details about the training can be found in the appendix section.
For an optimal $\vec{\theta}^*$ where the loss function converges to its minimum the quantum circuit is trained and can be used for classifying unseen data $\vec{x}$
\begin{align} \hat{y}=f(\vec{x};\vec{\theta}^*)=\text{arg}\max_{k}\bra{\Vec{x}} U^\dagger(\vec{\theta}^*)\Pi_kU(\vec{\theta}^*)\ket{\Vec{x}}.
\label{vqc}
\end{align}
Our classifier is considered as a strong one if $\hat{y}{=}f(\vec{x};\vec{\theta}^*)$ assigns the correct class label to most of the  unseen data $\vec{x}$.

\begin{figure}
  \centering
  \includegraphics[width=0.45\textwidth]{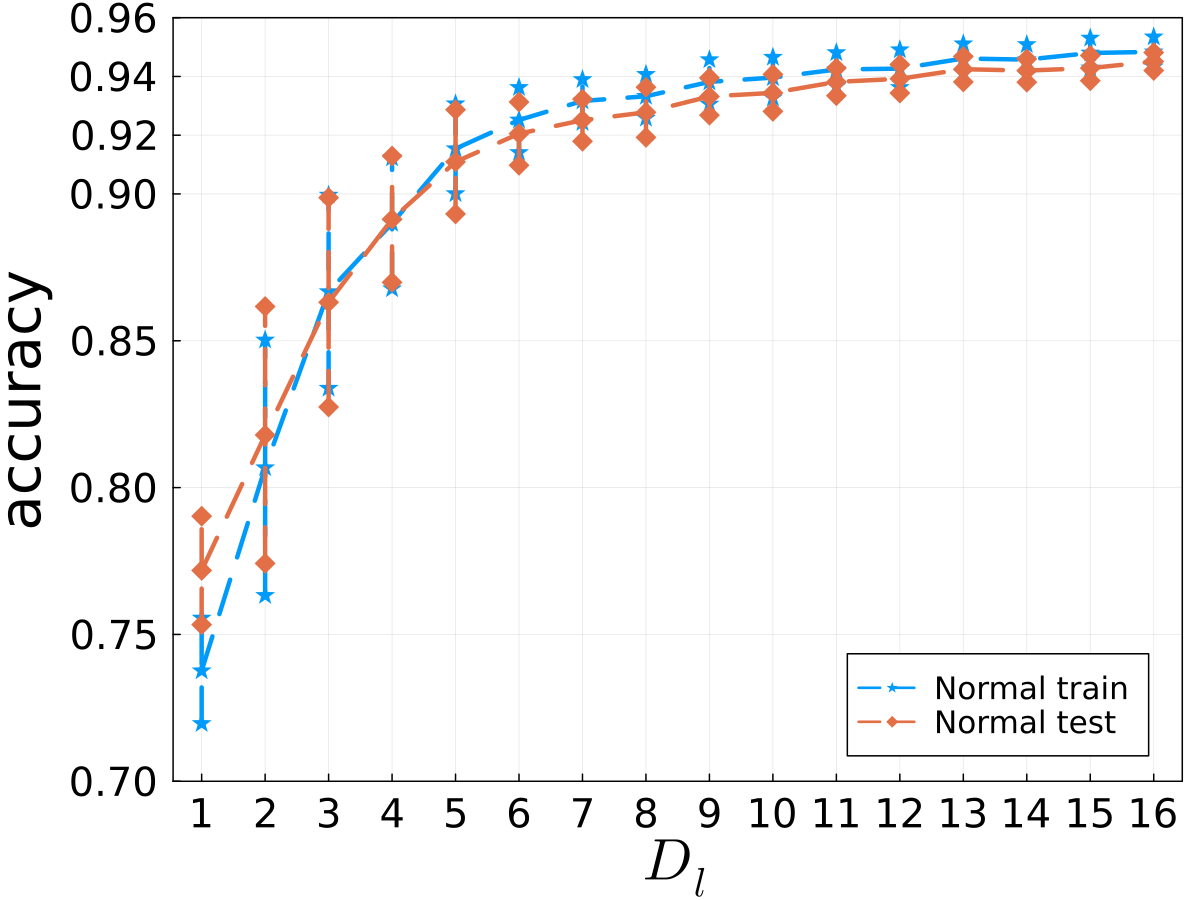}
  \caption[]{\textbf{Increasing layers in noise-free quantum circuits.} The training and testing accuracies of normal VQC for classifying MNIST dataset with odd digits $\{1,3,5,7\}$ are shown as a function of the depth $D_l$ of the parameterized circuit.  Each of the data points is averaged over $50$ different random initial sets of parameters and the error bars represent their standard deviation. The performance monotonically improves by increasing the circuit layers in a noise-free quantum computer. The closeness of the two types of accuracies shows that the training does not impose over-fitting.}
  \label{Fig.norm_amp}
\end{figure}
\subsection{The performance of VQC for Classical Datasets}
In order to show the performance of VQC for classifying classical data, we consider the MNIST dataset which contains handwriting digital images with $8 {\times 8}$ pixels (i.e. $N_\text{f}{=}64$ features)~\cite{Dua:2019}. Each pixel takes a number between 0 (perfectly white) to 1 (perfectly black). For the sake of simplicity and without loss of generality we only consider odd numbers and thus our classification has four different classes, labeled by digits 
$1$, $3$, $5$ and $7$. A typical image for each of these four classes is presented in Fig.~\ref{Fig.circuits}(b). The dataset contains $2267$ samples from which $1541$ samples are used for training and the $726$ unseen samples are used for testing the accuracy.
We train the circuit shown in Fig.~\ref{Fig.circuits}(a) with $N_\text{q}{=}6$ for various layers $D_l$. In Fig.~\ref{Fig.norm_amp} we plot both the training and test accuracies as a function of circuit layers $D_l$. 
Each data point is averaged over $50$ random initialization of the circuit parameters. 
The error bars show the variation of accuracy across these $50$ repetitions.  
Furthermore, both training and test accuracies are very close to each other which shows that the training is not affected by overfitting. 
Due to this, in what follows, we only report test accuracy as a quantification measure for the quality of our procedure. 
In addition, the accuracy is improved rapidly up to $D_l{\sim} 7$ layers before entering a slow convergence regime. 
In fact, one needs $D_l{=}12$ layers to achieve an accuracy of $0.94$ and even for up to $D_l{=}16$ one cannot still reach an accuracy of $0.95$. By increasing the number of layers the error bars decrease indicating robustness against parameter initialization.
Note that our quantum circuit is noise-free and all quantum gates operate perfectly. 
That is why the accuracy keeps improving by increasing the layers. 
In practice, since gates are imperfect and each of them induces noise in the system the accuracy has a more complex dependence on the circuit depth as will be discussed in the following sections.  
   
\subsection{VQC with ZNE}
NISQ quantum computers suffer from gate operations and short qubit coherence times. 
While single-qubit operations can be achieved with fidelity $\sim 0.999$~\cite{PhysRevLett.129.010502}, the two-qubit gates are more susceptible to noise. 
For the sake of simplicity, in order to simulate the effect of noise in NISQ computers one can consider two-qubit gates as the only source of noise in the system. 
In this paper, we emulate the effect of noise as a depolarizing channel which affects the operation of controlled-not gates on qubits $q$ and $q+1$ as  
\begin{equation}\label{eq:CNOT_Noisy}
 \xi_{CX}(\rho^{(q,q+1)})=\frac{P}{4}I^{(q,q+1)}+(1{-}P)U^{(q,q+1)}_{CX}\rho U^{(q,q+1)\dagger}_{CX},
\end{equation}
where $P$ quantifies the strength of decoherence. 
Note that this noise model is very pessimistic as the output is considered to be a maximally mixed state with probability $P$ which means that decoherence kills all the information in the system. 
In order to reduce the impact of noise in near-term quantum computers, error mitigation techniques~\cite{PhysRevX.7.021050,temme_error_2017,kandala_error_2019,endo_hybrid_2021,qin2022overview,cai2022quantum} have been developed for post-processing the noisy data. 
In this paper, we use the ZNE method~\cite{PhysRevX.7.021050,temme_error_2017,kandala_error_2019} to mitigate errors in primitive VQCs in which the zero-noise expectation value of an observable is extrapolated from its values at different noise levels. 
To achieve this, one has to systematically increase the noise in the system and measure the expectation value of the desired observable at different noise strengths. 
In our case, since the noise is assumed to be only in controlled-not gates, we can increase the noise strength by gate folding \cite{giurgica2020digital}: We replace each controlled-not gate with an odd number of controlled-not gates. 
Since $(U_{CX}^{(q,q+1)})^2{=}I^{(q,q+1)}$, then all odd powers of $U_{CX}^{(q,q+1)}$ is expected to be the same as one controlled-not gate. 
However, for the noisy operation $\xi_{CX}$ in Eq.~(\ref{eq:CNOT_Noisy}), the multiplication of controlled-not gates induces more noise in the system. 
We perform ZNE for circuits in which every controlled-not gate is replaced with $1$, $3$, $5$, and $7$ consecutive gates which approximately correspond to noise strengths of $P$, $3P$, $5P$, and $7P$, respectively. 
The error-mitigated result for $P{=}0$ is estimated through third-order polynomial extrapolation. 

\section{Ensemble-learning Algorithms}
Ensemble-learning classifiers have been introduced in classical machine learning literature for enhancing the precision of weak classifiers~\cite{ensemble_learning_review}. In these methods, a group of weak classifiers is combined to make a strong classifier with high accuracy. 
There are several ensemble-learning techniques for classification problems. The most prominent of such algorithms include  Bagging~\cite{breiman_bagging_1996} and AdaBoost~\cite{freund_desicion-theoretic_1995,hastie_multi-class_2009}.

The ZNE method, at best, remove the effect of noise in VQCs, and they usually cannot outperform noise-free quantum computers. Therefore, when error-free shallow circuits are insufficient for accurate classification, the improvement by ZNE is limited. 
In the following, we adopt two ensemble-learning error mitigation algorithms, namely Bagging and AdaBoost, for VQCs and show how these methods can enhance our classification accuracy. 

\subsection{Ensemble-Learning: Bagging VQC}
The Bagging algorithm has been developed as one of the most successful ensemble-learning techniques in the context of classical machine learning~\cite{breiman_bagging_1996}.
In the Bagging algorithm, a group of classifiers, each trained independently with different  training dataset $\mathcal{D}_i$, are combined to make a stronger one. After training, for any given data, all classifiers assign a class label, and the final prediction is decided by a majority vote among all these results. The simplicity of the Bagging algorithm has made it one of the most popular algorithms in classification problems.
Here, we show how a Bagging algorithm can be adapted for VQCs. To implement this, we train $L_{\text{c}}$ different VQCs with shallow circuits, all with equal layers. The difference between these classifiers is in the initialization of the parameters, which results in different optimal values of $\vec{\theta}^*$. Hence, one gets $L_{\text{c}}$ different VQCs, all trained independently. For unknown data $\vec{x}$, we use the majority vote among these $L_{\text{c}}$ classifiers to assign a class label. Therefore, the final classifier can be described as 
\begin{equation}
    \hat{y}=F_{\text{BG}}(\vec{x};\Vec{\theta}^*) = \text{arg}\max_{k}\sum_{l=1}^{L_\text{c}}\mathbb{I}(f_l(\vec{x};\Vec{\theta}_l^*)=k),
    \label{BG_Final}
\end{equation}
where $f_l(\vec{x};\Vec{\theta}_l^*)$ is a VQC, given in Eq.~(\ref{vqc}).

To see the performance of the Bagging algorithm, in Fig.~\ref{Fig.amp}(a) we plot the test accuracy as a function of $L_{\text{c}}$ for two types of noise-free circuits with $D_l{=}2$ and $D_l{=}3$ layers, respectively.  
Each data point is again averaged over $50$ random initializations.
As expected, in the absence of noise, the performance of the quantum circuit with $D_l{=}3$ layers always outperforms the circuit with $D_l{=}2$ layers. More importantly, even for such shallow circuits, the accuracy enhances by increasing the number of classifiers $L_\text{c}$ such that for $L_{\text{c}}{=}10$ one can achieve the accuracy of $0.8856$ (for the circuit with $D_l{=}2$ layers) and $0.9173$ (for the circuit with $D_l{=}3$ layers). To achieve a similar accuracy on a single circuit one needs a quantum circuit with $D_l{=}6$ layers ($0.9205$), see Fig.\ref{Fig.norm_amp}. Note that these are all for noise-free computers (i.e. perfect controlled-not gates with $P{=}0$) and as we will see later the improvement achieved by Bagging becomes even more pronounced in the presence of noise.
\begin{figure*}
  \centering
  \includegraphics[width=0.9\textwidth]{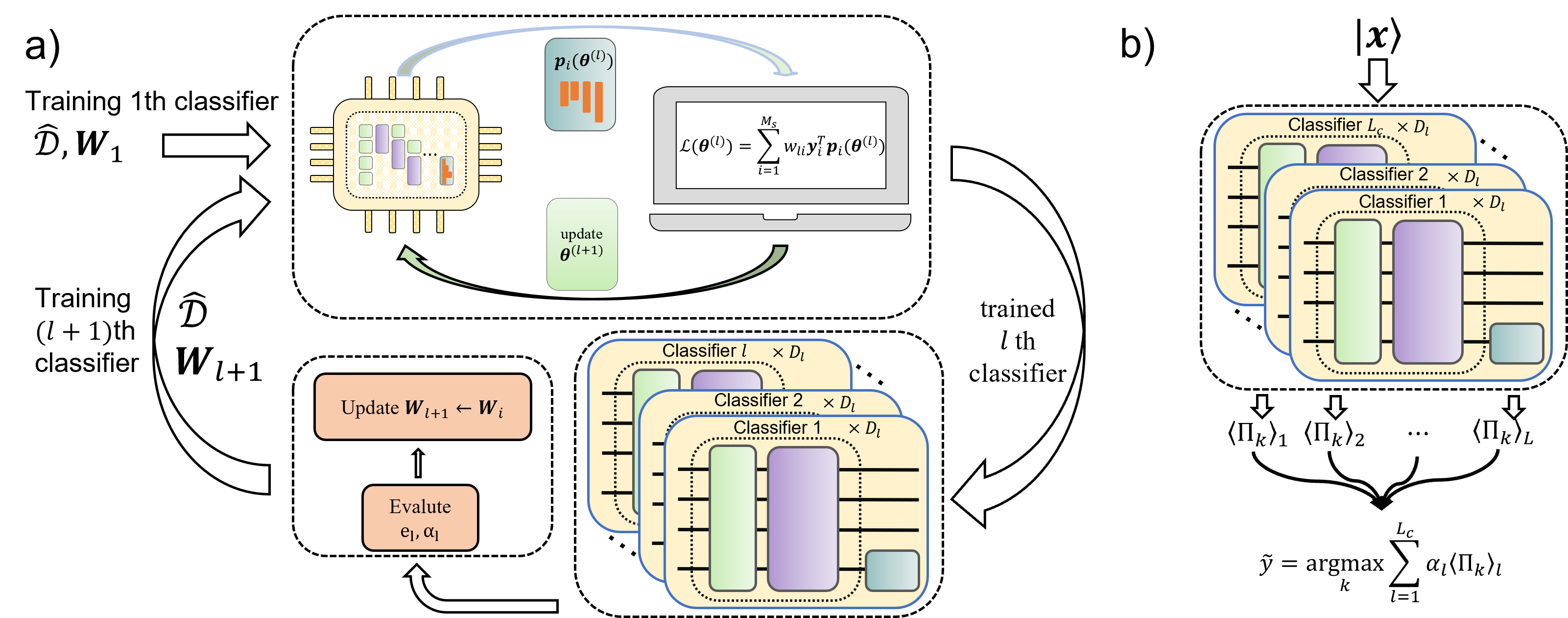}
  \caption{  (a) \textbf{The training process of AdaBoost VQC.} 
  The dataset $\hat{\mathcal{D}}$ and the data weights $\vec{W}_l$, which is initially taken to be uniform for the first weak VQC, are used to train the $l-$th weak VQC using a shallow circuit, shown in the top dotted box.  
  After training, the error rate $e_l$ is computed with which one can get the classifier's weight $\alpha_l$. Then the data weights are updated to get $\vec{W}_{l+1}$ using $\alpha_l$ and the previous data weights $\vec{W}_{l}$. The process repeats until all the $L_{\text{c}}$ classifiers are trained. 
  These trained weak VQCs are combined according to their weights $\alpha_l$ to make a single strong AdaBoost VQC with high accuracy.  
  (b) The unknown input data $\ket{\vec{x}}$ is fed into $L_\text{c}$ different trained weak VQCs.
  Then the probabilities $\langle\Pi_k\rangle_l$ of measurement outcome $k$ from different weak VQCs are averaged with weights $\alpha_l$. The final predicted class label is the $k$ with the largest outcome, namely $\text{arg}\max_k\sum_{l=1}^{L_\text{c}}\alpha_l\langle\Pi_k\rangle_l$.}
  \label{ada_flow}
\end{figure*}

\subsection{Ensemble-Learning: AdaBoost VQC}
AdaBoost  is an alternative ensemble-learning algorithm that is used to improve the accuracy of weak classifiers~\cite{freund_desicion-theoretic_1995,hastie_multi-class_2009}. It can be used for those classifiers that slightly outperform  a random guess, namely $0{\le} e {\le} (K_\text{c}{-}1)/K_\text{c}$~\cite{hastie_multi-class_2009}. While in the Bagging approach, the VQCs are trained in parallel (i.e. independently), in the AdaBoost scheme the VQCs should be trained sequentially.  
 We consider $L_\text{c}$ different quantum classifiers $f_l(\vec{x};\vec{\theta}_l)$, with $l{=}1,2,\cdots, L_\text{c}$. In the AdaBoost training process, one assigns a proper weight to each input data $\vec{x}_i$ in the loss function. After training each classifier (namely finding an optimal set  of parameters $\Vec{\theta}^*$), the weights are updated for training the next one, based on the performance of the last classifier. Hence, the training procedure for the classifiers is interconnected and can only be accomplished sequentially. 
 For simplicity, we assume that these quantum classifiers have the same circuit design with equal depths. However, each classifier starts with a different random initial parameterization $\vec{\theta}_l$ and uses a different loss function, depending on the weights. The AdaBoost algorithm pursues the following steps to make a single strong classifier via a proper interconnected training method of these $L_\text{c}$ classifiers:

\begin{itemize}
    \item \textbf{Step 1: Initializing the input weights.} We assign an initial weight $\vec{W}_1{=}[w_{1,1},w_{1,2}, \cdots,w_{1,M_\text{s}} ]$ with $w_{1,i}{=}1/M_\text{s}$ to all the inputs in the dataset. 

     \item \textbf{Step 2: Training the quantum classifier $f_l(\vec{x};\vec{\theta}_l)$.} We train the quantum circuit of the classifier $f_l(\vec{x};\vec{\theta}_l)$ (initially we start with $l{=}1$) using $\vec{W}_l$ and the loss function
\begin{equation}
 \mathcal{L}_l(\vec{\theta}_l)=-\sum_{i=1}^{M_\text{s}}w_{l,i}\Vec{y}_i^\mathsf{T}\log(\Vec{p}_i),
\end{equation}
When training finishes one gets the optimal parameter $\vec{\theta}_l^*$ which corresponds to the classifier $f_l(\vec{x};\vec{\theta}_l^*)$.

\item \textbf{Step 3: Computing error rate.} For the trained classifier $f_l(\vec{x};\vec{\theta}_l^*)$ one can compute the error rate as
\begin{equation}
\label{Eq.error_rate}
e_l=\sum_{i=1}^{M_\text{s}} w_{l,i}\mathbb{I}(f_l(\vec{x}_i;\vec{\theta}_l^*)\neq y_i).  
\end{equation}

\item \textbf{Step 4: Computing the classifier's weight.} Based on the error rate $e_l$, we can assign a weight to the classifier as
\begin{equation}
    \alpha_l=\log(\frac{1-e_l}{e_l})+\log(K_\text{c}-1)
\end{equation}
Note that for classifiers better than random guess, namely  $e_l<(K_\text{c}-1)/K_\text{c}$,  the coefficient $\alpha_l$ is always positive.

\item \textbf{Step 5: Updating the input weights.} For those input data $\vec{x}_i$that the classifier $f_l(\vec{x}_i;\Vec{\theta}_l^*)$ fails to estimate the correct class label $y_i$, we increase the input weight $w_{l+1,i}$. 
The reason is that during the training of the next classifier, this input data will have more impact on the loss function and thus might be correctly classified by the next classifier. 
The input weights are updated as  
\begin{equation}
    w_{l+1,i}=\frac{w_{l,i}}{Z_l}e^{\alpha_l\mathbb{I}(f_l(\vec{x}_i;\vec{\theta}_l^*)\neq y_i)},
\end{equation}
where $Z_l$ is the normalizing factor 
\begin{equation}
Z_l=\sum_{i=1}^{M_\text{s}}w_{l,i}e^{\alpha_l\mathbb{I}f_l(\vec{x}_i;\Vec{\theta}_l^*)\neq y_i)}.
\label{Z_factor}
\end{equation}

\item \textbf{Step 6: Training the next classifier.} Repeat from Step 2 until all the $L_\text{c}$ classifiers are trained. 

\item \textbf{Step 7: Combining the classifiers.} One can combine the trained classifiers in order to obtain a stronger one. The combination is weighted according to the strength of each classifier, quantified by $\alpha_l$:
\begin{equation}
    \hat{y}=F_{\text{AB}}(\vec{x};\Vec{\theta}^*) = \text{arg}\max_{k}\sum_{l=1}^{L_\text{c}}\alpha_l\mathbb{I}(f_l(\vec{x};\Vec{\theta}_l^*)=k).
    \label{AB_Final}
\end{equation}
   
\end{itemize}
The above steps are summarized in Fig.~\ref{ada_flow}(a). Note that the weak performance of the $L_\text{c}$ chosen classifiers can be due to different reasons such as shallow circuits or noisy gate operations.

\begin{figure*}
  \centering \includegraphics[width=0.9\textwidth]{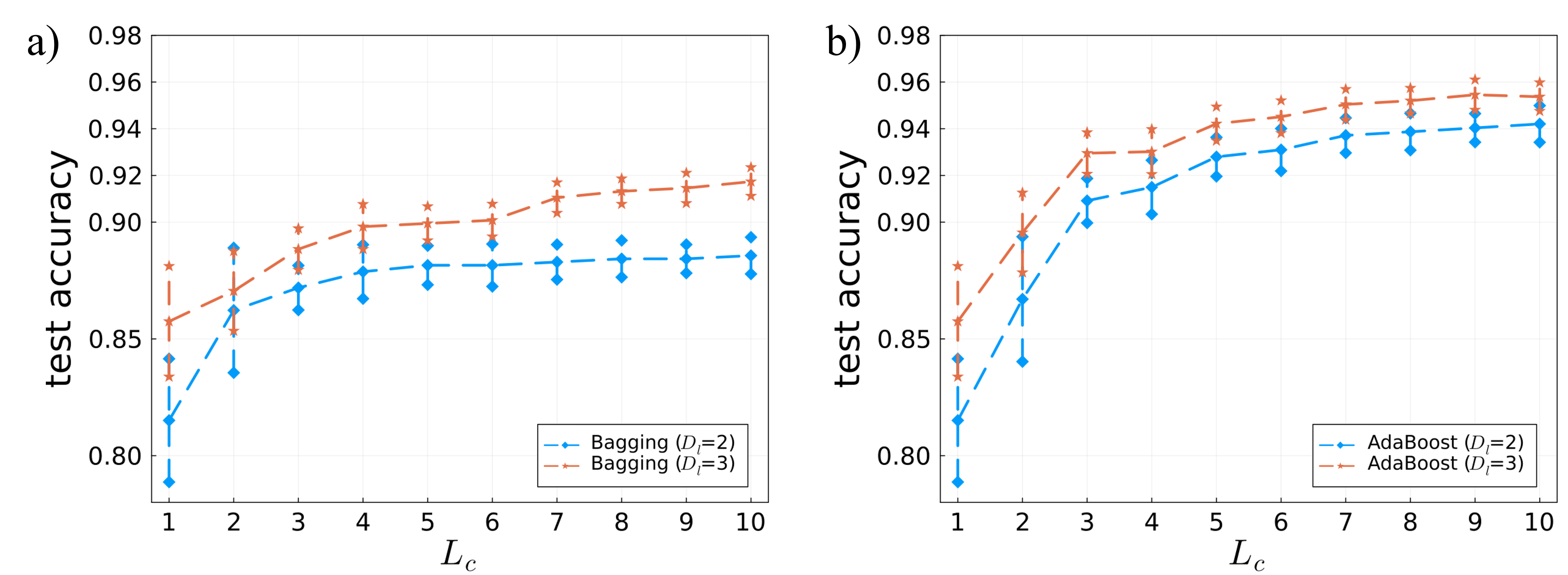}
  \caption{ \textbf{The performance of ensemble-learning VQCs in noise-free circuits. } The test accuracy of MNIST $\{1,3,5,7\}$ classification using ensemble-learning VQCs is plotted as a function of the number of weak classifiers $L_\text{c}$. The ensembles contain noise-free shallow quantum circuits with either  $D_l{=}2$ or $D_l{=}3$ layers. The panels represent: (a) Bagging VQC; and (b) AdaBoost VQC. For both algorithms, each data point is averaged over $20$ different samples of the initial parameters and the error bars are the standard deviation of those results. Increasing the number of classifiers monotonically enhances the performance of both algorithms.  
  Since the quantum circuits are noise-free, the performance of the circuit with $D_l{=}3$ layers is always better than the circuit with $D_l{=}2$ layers. 
   It is worth noting that for the same circuit layers $D_l$ and the number of classifiers $L_\text{c}$, the  AdaBoost VQC always outperforms the Bagging VQC. }
  \label{Fig.amp}
\end{figure*}

To see the performance of AdaBoost VQC, we first consider shallow VQC circuits whose gate operations are perfect (i.e. the controlled-not gates are noise-free with $P{=}0$) and thus their accuracy is only affected by the depth of their circuit. 
In Fig.~\ref{Fig.amp}(b) we plot the test accuracy as a function of $L_\text{c}$ for two types of circuits with only $D_l{=}2$ and $D_l{=}3$ layers. 
As the figure shows, by increasing the number of classifiers the accuracy increases. In addition, $3$-layer circuits provide better accuracy in comparison with the $2$-layers circuits. 
This is  because, in the absence of noise, a $3$-layer circuit naturally performs better than a $2$-layer one. One can compare the performance of  Bagging and  AdaBoost in Figs.~\ref{Fig.amp}(a) and (b) when the circuit depths are the same. The figures clearly show that AdaBoost can outperform Bagging.  For instance, by considering circuits with $D_l{=}3$ layers, the AdaBoost VQC  with  $L_\text{c}{=}6$ classifiers can achieve an accuracy of $0.95$ while the Bagging VQC even with $L_\text{c}{=}10$ classifiers cannot exceed $0.92$ accuracy. This is because, during the AdaBoost sequential training, each classifier is provoked to correct the mistakes of the previous classifier through weight updating. In contrast, in the Bagging algorithm, the classifiers are trained in parallel and independent from each other. Therefore, the mistakes are not  corrected as efficiently as in the AdaBoost algorithm.     

\begin{figure*}
  \centering
   \includegraphics[width=1\textwidth]{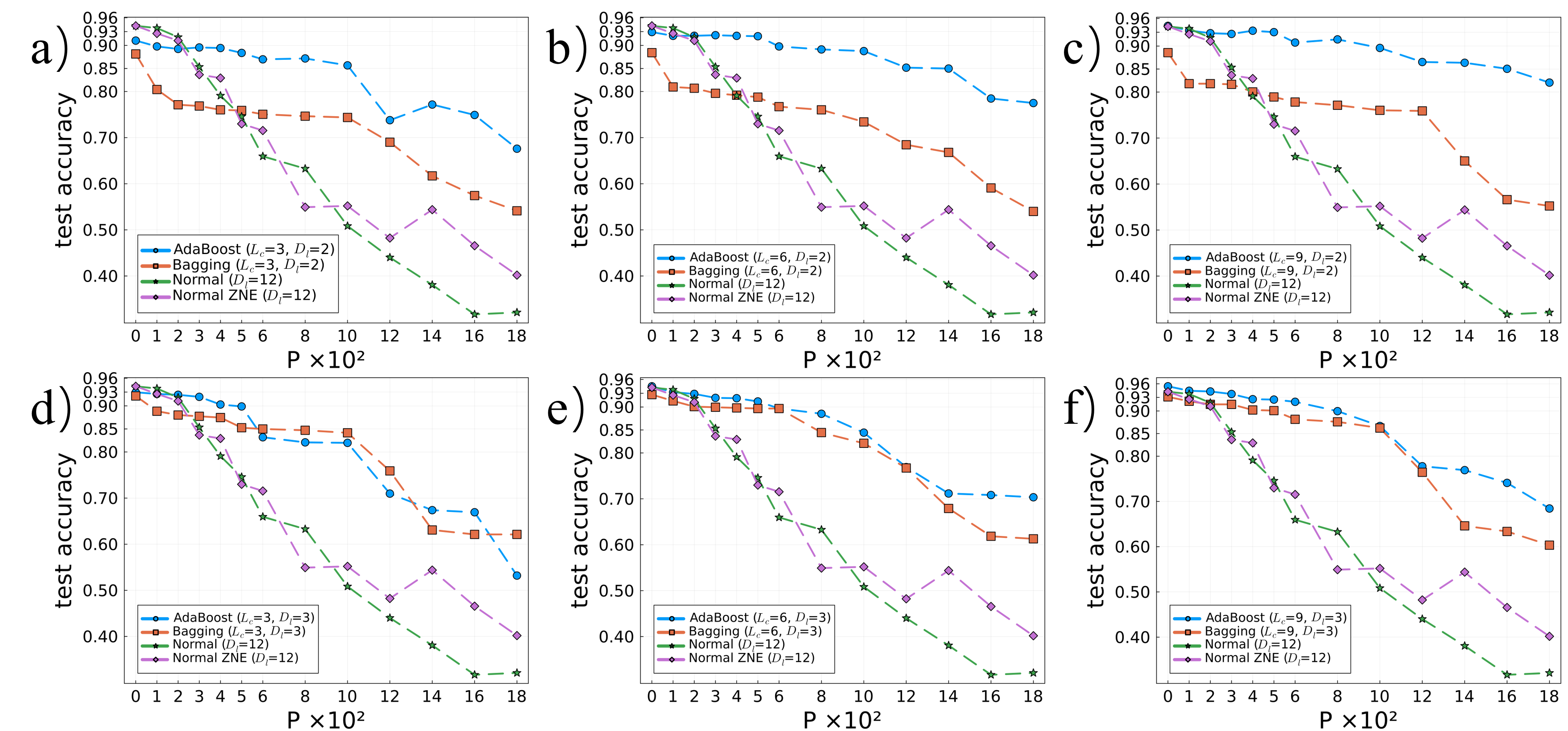}
  \caption{ \textbf{Comparison of various VQC algorithms on noisy quantum circuits.} The test accuracy of four different VQC algorithms is plotted as a function of decoherence rate $P$. The four strategies include normal VQCs, performed on a deep circuit of $D_l{=}12$ layers, with and without ZNE, as well as our ensemble-learning algorithms, namely Bagging VQC and AdaBoost VQC. In the upper panels, both Bagging and Adaboost are performed on shallow circuits with $D_l{=}2$ layers and ensembles of size: (a) $L_\text{c}{=}3$; (b) $L_\text{c}{=}6$; and (c) $L_\text{c}{=}9$ classifiers, respectively. In the lower panels,  both Bagging and Adaboost are performed on shallow circuits with $D_l{=}3$ layers and ensembles of size: (d) $L_\text{c}{=}3$; (e) $L_\text{c}{=}6$; and (f) $L_\text{c}{=}9$ classifiers, respectively. Each of the data points plotted in these panels is averaged over $50$ random samples of initial parameters. The results show that while conventional error mitigation ZNE can indeed enhance the classification accuracy of deep circuits, its performance remains below ensemble-learning classifiers with shallow circuits, as $P$ increases. The best outcome is indeed achieved by AdaBoost whose performance significantly enhances as the number of classifiers increases and remains very robust even at large decoherence rates. 
}
  \label{Fig.Amp_P}
\end{figure*}

\subsection{Ensemble-Learning VQCs on NISQ computers}
In this section, we consider noisy quantum computers in which controlled-not gates are noisy and operate according to Eq.~(\ref{eq:CNOT_Noisy}). 
The strength of noise is quantified with decoherence rate $P$, which affects all the two-qubit gates of the circuit equally. 
We compare four different scenarios: (i) a normal VQC with a deep circuit of $D_l{=}12$ layers without ZNE; (ii) a VQC with a deep circuit of $D_l{=}12$ layers with ZNE; (iii) Bagging VQC with shallow circuits of $D_l{=}2$ and $D_l{=}3$ layers with various numbers of classifiers; and (iv) AdaBoost VQC with shallow circuits of $D_l{=}2$ and $D_l{=}3$ layers with various numbers of classifiers. 
We first fix the circuit layer $D_l{=}2$ for Bagging and AdaBoost and plot the test accuracy as a function of decoherence rate $P$ in Figs.~\ref{Fig.Amp_P}(a)-(c), for $L_{\text{c}}{=}3$, $6$ and $9$ classifiers, respectively. 
As the figure shows, the test accuracy for a normal VQC with a deep circuit of $D_l{=}12$ decays rapidly as $P$ increases. ZNE can indeed enhance the accuracy for such a deep circuit, but for larger $P$ the decay is still significant. 
Interestingly, a shallow Bagging VQC with $D_l{=}2$ can outperform the deep circuit classifier even with ZNE when $P>0.06$. 
Increasing the number of classifiers from $L_{\text{c}}{=}3$ to $L_{\text{c}}{=}9$ slightly improves the performance of Bagging. 
Remarkably, the AdaBoost algorithm with even shallow circuits of $D_l{=}2$ layers can outperform the other scenarios for noise rates of $P>0.02$ and remains stably high even for very strong decoherence rates up to $P{=}0.18$.   

Similarly, one can consider the Bagging and the AdaBoost with $D_l{=}3$ layers in Figs.~\ref{Fig.Amp_P}(d)-(f) for $L_{\text{c}}{=}3$, $6$ and $9$ classifiers, respectively. 
In this case, the Bagging and AdaBoost outperform deep circuits with ZNE when $P>0.02$. As $P$ increases, the performance of Bagging and AdaBoost remains fairly close to each other for $L_{\text{c}}{=}3$ and $L_{\text{c}}{=}6$ classifiers. 
By increasing the number of classifiers $L_{\text{c}}$ or noise rate $P$, again AdaBoost outperforms Bagging. 
Note that the AdaBoost algorithm is hugely benefited by increasing the number of classifiers due to its interconnected training method, which improves the classifiers based on the mistakes of the previous ones. 
Another interesting observation is that in very noisy quantum computers, i.e. large $P$, AdaBoost with $D_l{=}2$ layers is better than AdaBoost with $D_l{=}3$ layers. 
This is because deeper circuits naturally have more two-qubit gates and thus are more susceptible to the effect of noise. 

In summary, both our ensemble-learning error mitigation methods, namely Bagging VQC and AdaBoost VQC, provide a significant improvement over conventional ZNE method. 
This is a general behavior and can also be observed for rotation encoding of the input data (results not shown). 
Moreover, thanks to its interconnected training method, the AdaBoost algorithm can outperform the Bagging, in particular, when the number of classifiers increases.
The AdaBoost accuracy enhancement over the other methods becomes even more pronounced when the quantum computer is subjected to strong decoherence, namely large $P$.

\begin{figure*}
  \centering \includegraphics[width=0.95\textwidth]{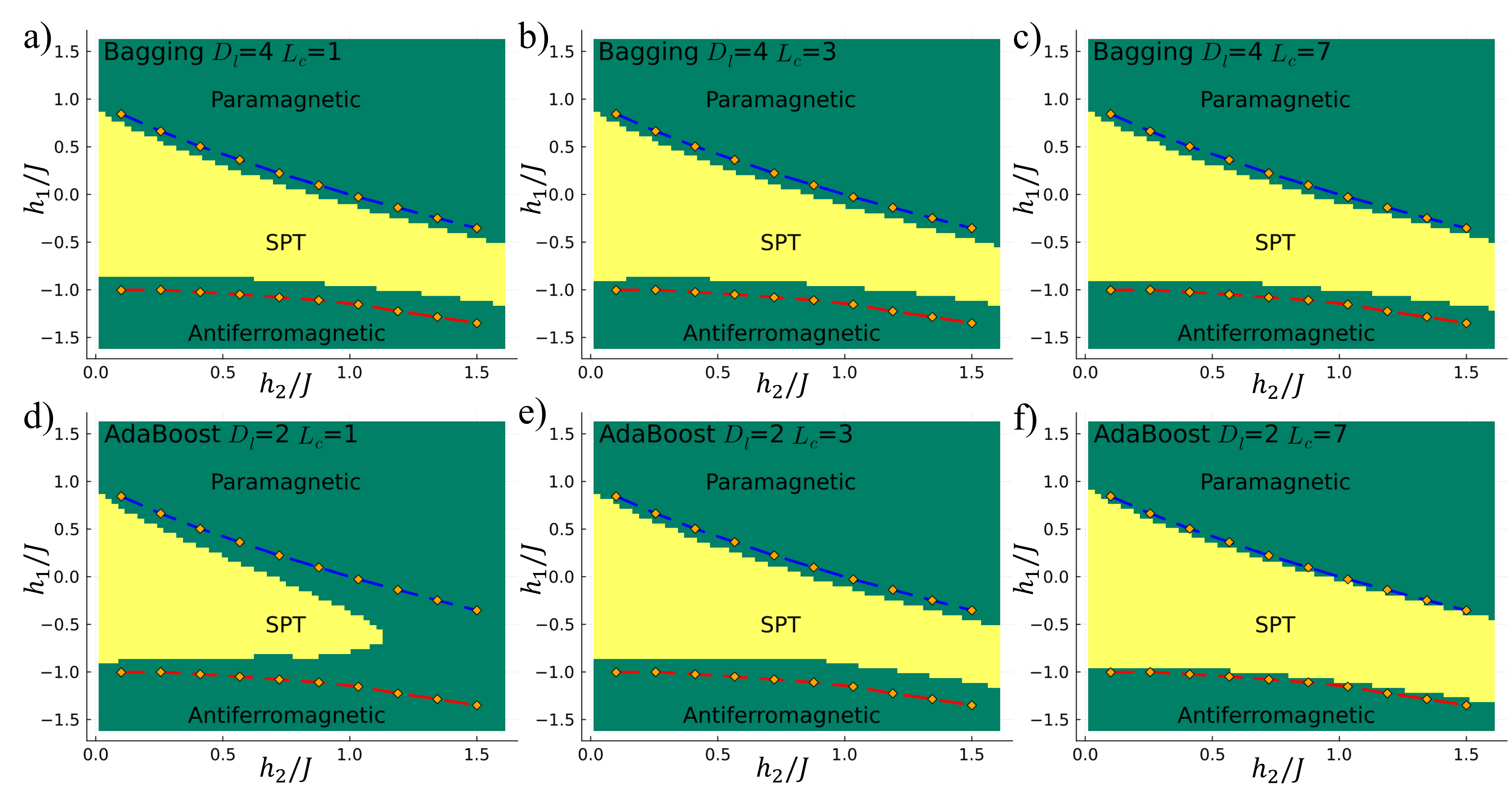}
  \caption{ \textbf{VQC for quantum datasets. } The test performance for phase recognition of the ground state of the SPT Hamiltonian~(\ref{Ham}) with $15$ qubits as a function of $h_1/J$ and $h_2/J$. The upper panels show the performance of Bagging for quantum circuits with $D_l{=}4$ layers with ensembles of size: (a)  $L_\text{c}{=}1$; (b) $L_\text{c}{=}3$; and (c) $L_\text{c}{=}7$ classifiers, respectively. The lower panels show the performance of AdaBoost for quantum circuits with only $D_l{=}2$ layers with ensembles of size: (d)  $L_\text{c}{=}1$; (e) $L_\text{c}{=}3$; and (f) $L_\text{c}{=}7$ classifiers, respectively. The blue and red lines represent the real phase boundaries computed through density matrix renormalization group~\cite{mcculloch2008infinite,cong_quantum_2019}. Note that using a single circuit $L_\text{c}{=}1$ is not really an ensemble learning but we just include it to show how the results improve as the number of classifiers increases. } 
  \label{Fig.SPT}
\end{figure*}

\subsection{Classification of Quantum Data}
In this section, we apply our ensemble classification methods  to a quantum dataset. The input data are quantum states which are taken from the ground state of a chain of $N_\text{q}$ qubits interacting via Hamiltonian
\begin{eqnarray}
    H=&-& J\sum_{i=1}^{N_\text{q}-2}\sigma_z^{(i)}\sigma_x^{(i+1)}\sigma_z^{(i+2)}-h_1\sum_{i=1}^{N_\text{q}-1}\sigma_x^{(i)}\sigma_x^{(i+1)}\cr &-& h_2\sum_{i=1}^{N_\text{q}}\sigma_x^{(i)},
    \label{Ham}
\end{eqnarray}
where  $J$ is the three-body spin coupling, $h_1$ is the two-body spin exchange interaction and $h_2$ is the magnetic field. Note that the three-body interaction term flips a central spin with the addition of a phase that depends on the quantum states of its neighbors. This Hamiltonian commutes with two string operators
\begin{equation}
X_{\text{odd (even)}}=\prod_{i\in\text{odd (even)}}\sigma_x^{(i)}. 
\end{equation}
This implies that the Hamiltonian has a $\mathbb{Z}_2\times\mathbb{Z}_2$ symmetry which results in the emergence of a Symmetry-Protected Topological (SPT) phase which is described by a non-local order parameter~\cite{PhysRevB.86.125441,PhysRevLett.109.050402}. In Ref.~\cite{cong_quantum_2019} the phase diagram of this Hamiltonian has been determined through density matrix renormalization group analysis~\cite{mcculloch2008infinite}. The Hamiltonian has three different phases as $(h_1/J,h_2/J)$ vary, namely antiferromagnetic, paramagnetic, and SPT phases. 
In the absence of two-body interaction, i.e.
$h_1{=}0$, the Hamiltonian becomes
solvable via Jordan–Wigner transformation and shows a quantum phase transition from the SPT to the paramagnetic phase at a specific value of $h_2/J$. Recently, the phase diagram of this system has also been determined through quantum convolution neural networks~\cite{cong_quantum_2019} which has been experimentally realized in superconducting quantum computers for a system of size $N_\text{q}{=}7$~\cite{beckey_variational_2022}. 

Here, we use our ensemble classification methods for determining the phase diagram of the system. The quantum circuit is exactly the same as before, shown in Fig.~\ref{Fig.circuits}(a), with one important difference. Since the input is itself a quantum state, the encoder is no longer needed and the quantum state can directly be fed into the parameterized circuit.
Similar to the approach of Ref.~\cite{cong_quantum_2019}, we only measure the last qubit despite having three phases, i.e. three classes. This method labels the phases as SPT and non-SPT phases. Since anti-ferromagnetic  and  paramagnetic phases are well separated and have no boundary they can be easily recognized in the phase diagram, as we will see in the following.

First, we focus on the Bagging algorithm for phase recognition of the SPT Hamiltonian with $N_q{=}15$ qubits using an ensemble of circuits with $D_l{=}4$ layers. We consider the phase diagram in the $(h_1/J,h_2/J)$ plane with the resolution of $64\times 64$ pixels. For training the circuit, We randomly select the ground states of $M_\text{s}{=}400$ random samples in the $(h_1/J,h_2/J)$ plane, as our training data. In Figs.~\ref{Fig.SPT}(a)-(c) we plot the result of our Bagging VQC for an ensemble of $L_\text{c}{=}1$, $3$, and $7$ classifiers, respectively. The phase boundaries, computed by density matrix renormalization group~\cite{mcculloch2008infinite,cong_quantum_2019}, are plotted by the blue and red lines.  
As evident in the figures, Bagging VQC can indeed capture the phase diagram and the precision becomes better as the number of classifiers increases. It is worth emphasizing that for shallower circuits with the depth $D_l<4$ layers, the precision for capturing the phase diagram goes down (results not shown). In particular, the performance is poor for circuits with $D_l{=}2$ layers, no matter how many classifiers we use. This shows that increasing the number of classifiers alone cannot compensate the circuit dept.  This is because the classifiers are trained independently and their weakness cannot be improved during training.

Second, we also exploit AdaBoost for capturing the phase diagram of the Hamiltonian with very shallow circuits of $D_l{=}2$ layers. Similar to the previous cases, we use the same circuit as shown in Fig.~\ref{Fig.circuits}(a) without the encoder part. 
We use the same dataset that we used for the Bagging algorithm.  
In Figs.~\ref{Fig.SPT}(d)-(f) we depict the phase diagram of the system using AdaBoost circuits with the depth of $D_l{=}2$ layers and $L_{\text{c}}{=}1$, $3$ and $7$ classifiers, respectively. Note that the AdaBoost can only become effective for more than one classifier. As the figures clearly show, the AdaBoost protocol can indeed determine the phase diagram even with shallow circuits with only $D_l{=}2$ layers. As expected, the precision is improved as the number of classifiers $L_{\text{c}}$ increases. 
In particular, for $L_{\text{c}}{=}7$ the phase boundaries between the SPT and the other phases are captured quite precisely. The fact that circuits with only $D_l{=}2$ layers are enough for recognizing the phase boundaries already shows the superiority of AdaBoost VQC over Bagging VQC. As mentioned before, this is because in AdaBoost VQC the training of classifiers is not independent of each other in such a way that each classifier tries to correct the errors of the previous ones through weight updating.

\section{CONCLUSIONS}
We introduced two ensemble-learning error mitigation algorithms, namely Bagging and AdaBoost, for VQCs. 
These algorithms can significantly enhance the precision of classification using only shallow quantum circuits with very few parameters to train. 
Our protocols have been tested on both classical (handwriting digits) and quantum (the phase recognition of an SPT Hamiltonian) datasets. 
Considering imperfect NISQ computers, our ensemble-learning error mitigation methods significantly outperform the ZNE method in classification tasks. 
Thanks to its interconnecting training approach, which tends to correct the mistakes of one classifier in the training of the next one, the AdaBoost method achieves better accuracy and shows better robustness against noise than the Bagging algorithm. 
The superiority of AdaBoost over ZNE and Bagging becomes even more prominent when the number of classifiers increases, in particular at the large noise limit.

Our ensemble-learning error mitigation techniques are very general. For classification problems, they are  applicable to both classical and quantum datasets and work for both amplitude and rotation encodings. The application of our ensemble-learning error mitigation methods is not limited to classification and can be generalized to other supervised machine-learning problems such as Kernel learning and regression. Moreover, it can also be used for non-variational classification methods such as quantum support vector machines. Our ensemble-learning VQCs are NISQ-friendly and very distinct from other ensemble-learning proposals~\cite{schuld_quantum_2018,wang_quantum_2020,arunachalam_quantum_2020,izdebski_improved_2020,macaluso2020quantum} which are hardware demanding, relying on multi-qubit controlled unitaries and quantum subroutines such as quantum phase estimation, Grover search and quantum mean estimation.\\ 
Therefore, our approach can solve a broad category of machine learning problems with today's NISQ technologies. 

\begin{acknowledgments}
The authors acknowledge support from the National Key R\&D Program of China (Grant No. 2018YFA0306703). A.B. thanks the National Natural Science Foundation of China (Grants No. 12050410253, No. 92065115, and No. 12274059), and the Ministry of Science and Technology of China (Grant No. QNJ2021167001L) for their support. X.W. thanks the National Natural Science Foundation of China (Grant No.~92265208) for their support. Y.L. thanks the National Natural Science Foundation of China (Grant No. 12225507 and No. 12088101) for their support. We also thank Guanyu Zhou for helpful discussions and Chu Guo for package ``VQC.jl''.
\end{acknowledgments}
~
\appendix*
\section{TRAINING THE QUANTUM CIRCUIT}
For numerical simulations, we rely on Julia packages ``VQC.jl" and ``QuantumCircuit.jl". In the training process of our VQCs, we use Adam optimizer~\cite{kingma_adam_2017}, which is a gradient-based  method, with a learning rate of $5\times 10^{-3}$, to update the quantum circuit parameters $\Vec{\theta}$. 
The gradients are obtained by Automatic differentiation methods supported by the VQC.jl package.
The optimization iterations of the training procedure are $500$ times for the weak quantum classifiers in both AdaBoost VQC and Bagging VQC and $1500$ times for the normal deep VQC.
In order to be initialization-independent, for noise-free and noisy circuits, the performance is averaged over $50$ and $20$  random initial samples, respectively. 

For quantum classification of the SPT Hamiltonian, the training dataset takes the ground state of $M_\text{s}{=}400$ random samples in the $(h_1/J,h_2/J)$ plane.
To show the performance of the classifier, we depict the phase diagram with the resolution of $64 \times 64$ averaged in the $(h_1/J,h_2/J)$ plane, as shown in Fig.~\ref{Fig.SPT}. 
The optimization iteration is fixed to $1000$ and each data point has been averaged over $10$ different random initial samples.

\bibliography{reference}
\end{document}